# But What Behavior?

Robert C. Froemke[1-4*]

[1] Department of Neuroscience, New York University Grossman School of Medicine, New York, NY 10016 USA.

[2] Department of Otolaryngology, New York University Grossman School of Medicine, New York, NY 10016, USA.

[3] Translational Neuroscience Institute, New York University Grossman School of Medicine, New York, NY 10016, USA.

[4] Center for Neural Science, New York University, New York, NY 10003, USA.

* Correspondence: robert.froemke@med.nyu.edu

Permanent address: 435 East 30th Street, Science Building 1215, New York, NY 10016

**Abstract:**

What is a ‘natural’ behavior? I argue that the study of natural behaviors is often the study of the spontaneous behaviors of animals placed in quantifiably different environments. For behavioral generalists such as rodents, humans, and many other species, there may be no such definable construct as a native habitat or ‘natural behavior’, due to their successful abilities and needs to rapidly adapt to a wide range of different ecosystems. Instead of prioritizing naturalness, it may be more essential to determine objective outcome measures related to specific behaviors; i.e., which sequences of behaviors and adaptive mechanisms allow animals to survive and reproduce, across a range of dynamic or hazardous physical and social environments.

**Main Text:**

In their influential review, Krakauer et al. [1] proposed that thorough behavioral analysis should be a pre-requisite for systems neuroscience and not an afterthought for trying to interpret the results of neurophysiological experiments. They provide four examples, each of which are major success stories in the field and continue to be fruitful lines of investigation today: bradykinesia in Parkinson's disease, sound localization in birds and mammals, the jamming avoidance response in weakly electric fish, and motor learning in primates. As the authors explained, the significance of these studies is due to computational theories that both led to and were refined by experimental testing. However, most behaviors performed by humans and other animals lack formal theoretical frameworks to guide investigation via quantifiable predictions and falsifiable hypothesis generation. How then should we choose behaviors to study?

**The case of sound source localization**

One of my favorite papers on sound localization is Hofman et al. [2], in which four adult human subjects wore ear molds that subtly changed the acoustic transfer function of the outer ear pinna (**Fig. 1A**). These were worn essentially continuously for a period of six weeks, without specific instruction or training, and sound localization abilities were measured in these subjects before, after, and occasionally throughout the six-week period. Sound localization specifically in elevation was at first greatly impaired, but gradually recovered over several weeks. This indicated that a form of online learning or adaptation occurred to reshape the neural representation of where sounds are in space, after changes in elevation cues from altering the pinna. When the mold was removed, there was no additional deficit, indicating that the neural circuits for sound localization support multiple mappings and instant re-mapping to a previously-learned set of spatial cues (**Fig. 1B**).

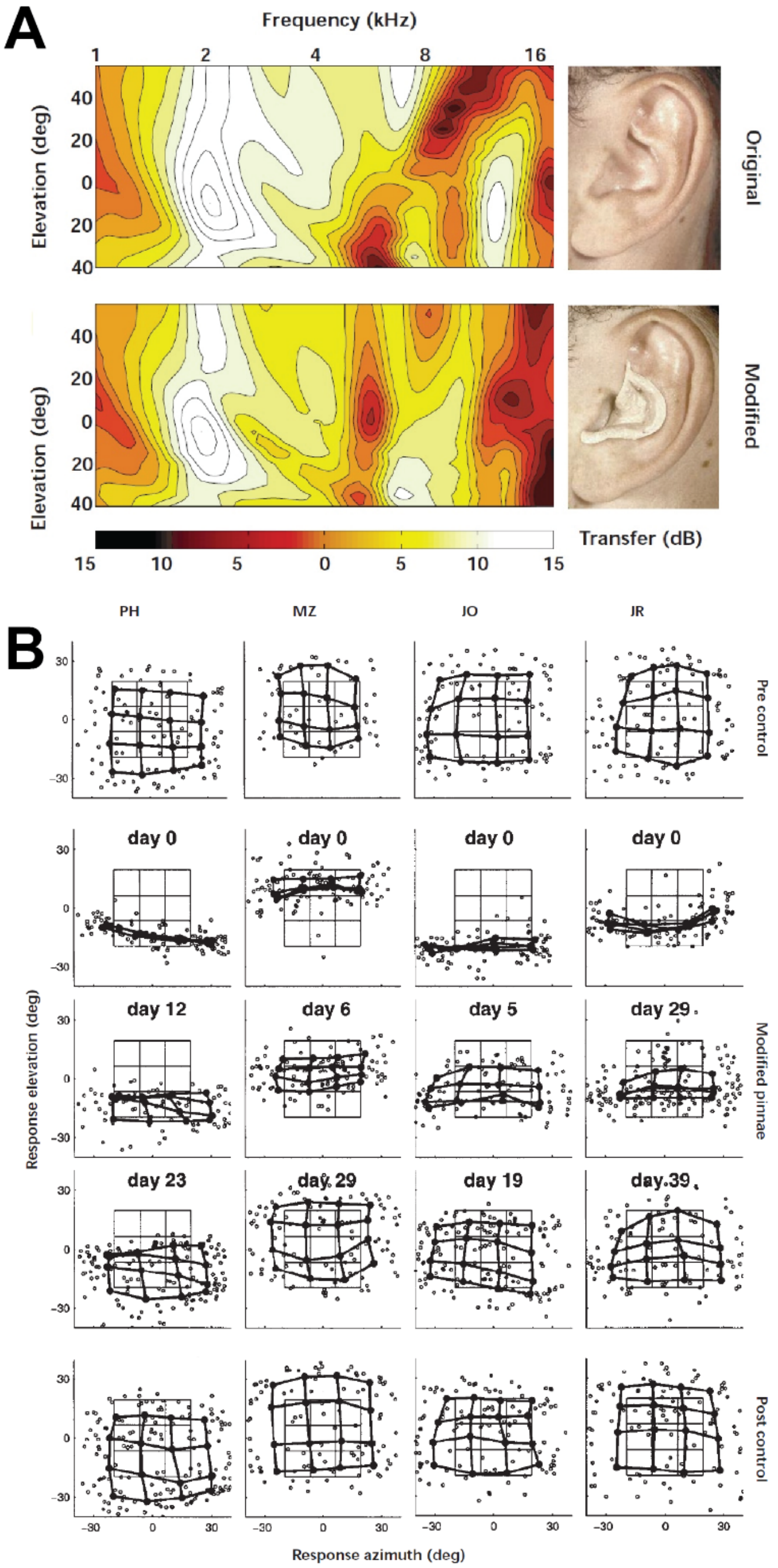

**Figure 1.** Sound localization plasticity induced with an artificial ear mold. **A**, Effect of molds on pinna acoustic transfer of one subject with normal ear (top) and wearing ear mold (bottom). Color, amplitude (in dB) of the transfer function of change to sound pressure amplitude of tones of different frequencies and elevations (lighter, amplification; darker, attenuation). **B**, Localization behavior of four subjects (one per column) before (first row), during (days 0+), and immediately after (last row) wearing the ear mold. Circles, individual saccades; thick lines, average saccades; thin lines, target matrix of sound source speakers. From Hofman et al. [2].

One remarkable aspect of this study is that the learning occurred ‘naturally’, i.e., it was wildly uncontrolled with different transfer functions and likely different acoustic experiences across each of the four subjects, so uncontrolled and individualized that the authors did not attempt to examine or analyze the experiences of each subject during this period. This study is important and powerful not because of that naturalistic aspect, but in spite of it: the neural mechanisms of sound localization plasticity are robust enough to be engaged despite the vagaries of personal experience, assessed with rigorous psychophysical testing in the lab, and produced by a fairly simplistic artificial manipulation (a silicone ear mold).

Thus Hofman et al. [2] made three discoveries: that sound localization abilities are plastic due to online and seemingly-unsupervised updating mechanisms (remaining still quite mysterious), that the pinna shape regulates localization in elevation but not azimuth, and that the human brain can simultaneously store multiple maps for sounds in space and switch between them when needed. These are interesting findings despite the lack of obvious ethological significance to humans for adult sound localization plasticity, and that the manipulation (an ear mold) and behavioral testing (saccades to a speaker in the lab) might not be considered ‘natural’. But what does this term ‘natural’ mean, and is it an important or useful concept for behavior and biology?

**Partner preference in voles**

One definition of natural (or at least naturalistic) behavior is just examining the spontaneous behaviors of animals in specific physical and social environments, outside the context of well-controlled conditioning studies or task performance. Here ‘spontaneous’ behavior means anything the animal does, not just those actions yoked to some feature of the experimental design. Perhaps exemplar of how social behaviors can best be studied comes from examining partner preference in prairie and montane voles. Early observations by Sue Carter, Lowell Getz, and colleagues found that when wild voles were trapped, prairie voles tended to be caught in pairs with the same pairs repeatedly caught, while nearby montane voles were usually caught in isolation (**Fig. 2A**, left). This suggested that prairie but not montane voles form enduring preferences for specific mating partners, which might relate to species differences or specializations of neural circuits and modulatory signals such as the central oxytocin system [3] [4]. To study the mechanisms of formation and maintenance of this complex process, it was necessary to develop a lab assay of partner preference (**Fig. 2A**, right). This led to many fruitful years of studies on the genetics, neuroendocrinology, and neural basis of partner preference and social behavior [5] [6].

In 2023 though, Berendzen et al. [7] published a landmark study in which they generated oxytocin receptor knockout voles. In contrast to a long literature demonstrating the importance of oxytocin signaling for social behavior and pair bond formation, the knockout voles were initially reported to have similar propensities for partner preference as measured with standard three-chamber assays (**Fig. 2B**). It is worth noting that the knockout vole mothers had substantially lower litter survival (~50% compared to ~100% for wild-type voles) likely due to nursing defects (**Fig. 2C**), the methods section reported significant maternal mortality specifically of the knockout voles, and a follow-up study from the same group [8]● reported that if assessed over a shorter time period there was a clear deficit in partner preference formation in the knockout voles (**Fig. 2B**).

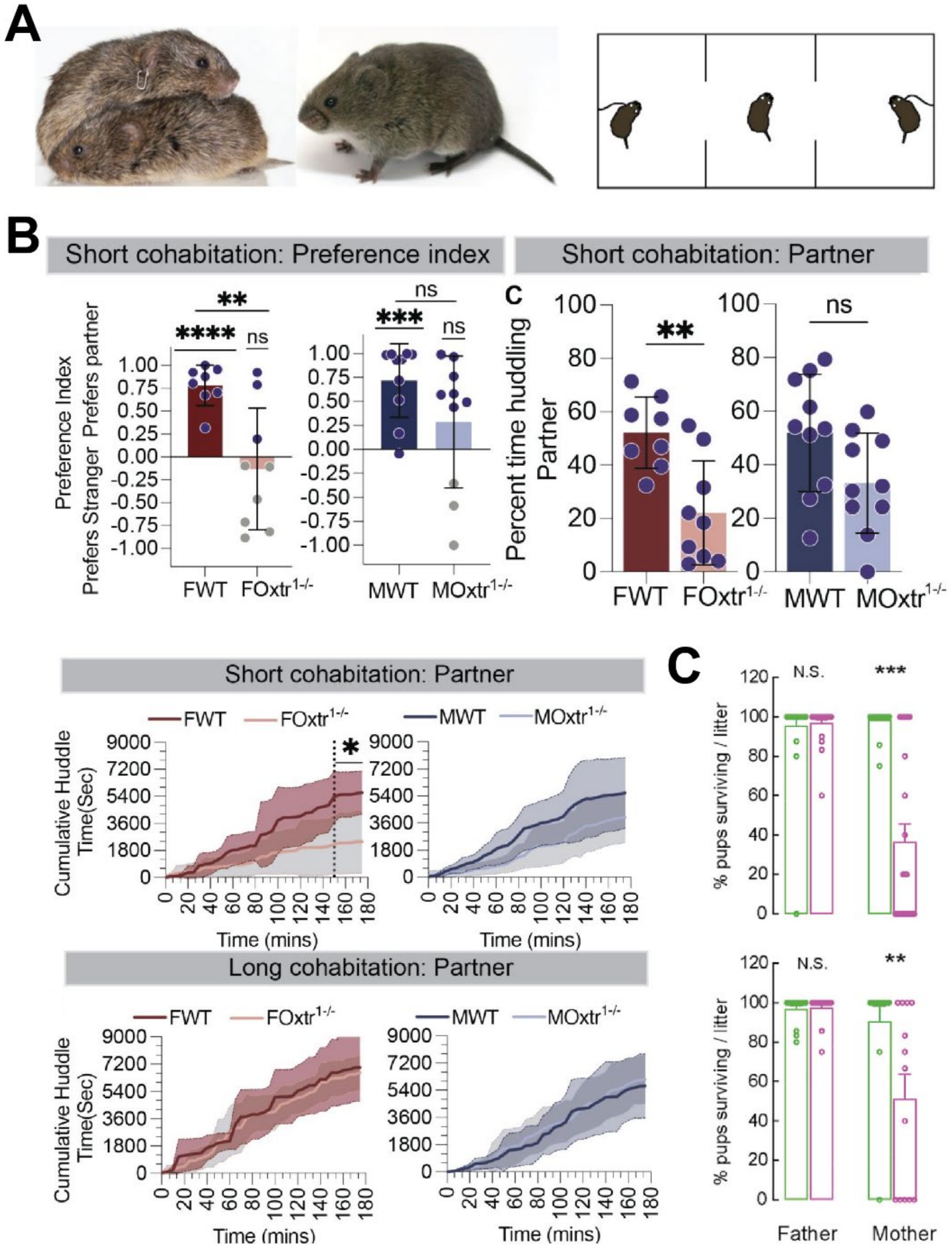

**Figure 2.** Behavioral assessment in wild-type and oxytocin receptor knockout prairie voles. **A**, Left, prairie vole partner preference measured by huddling; montane voles tend to be solitary. Right, three-chamber assay schematic for huddling quantification as proxy for partner preference formation/social bonding. **B**, Top, partner preference in wild-types but not oxytocin receptor knockout prairie voles after short (6 hr) cohabitation; bottom, partner preference in both after long (5 day) cohabitation. **C**, Extensive pup mortality born to receptor knockout females. Green, wild-type; red, knockouts. From [6] [7] [8]●.

This work did not test nor demonstrate that oxytocin receptors are dispensable for social bonding. Social bonding is a complicated neurobiological and psychological process that is ill-defined even in humans, much less non-human species. Partner preference assays (or variants thereof) are necessarily low-dimensional behavioral tests of the sort required for laboratory studies of complex processes, to scientifically examine one or a few aspects of social interactions in a replicable and time-limited manner. Instead, I see this as a reminder to be uncomfortable with the relation between complex behavioral phenomena of interest (here pair bonding, but also including disorders such as depression or schizophrenia), and models of those behaviors more amenable to experiment (e.g., a low dimensional partner preference assay). Even seemingly-straightforward measures such as partner preference involve a number of nuanced interactions that depend on context and testing conditions. Models will ultimately always be disappointing once their limits have been reached, and the main goal of experimentation is to test those limits and determine where models need revisions (such as generation of knockout voles by the Manoli lab).

I find it encouraging that the vole community continues to lead by example and work to develop other systems scaling up complexity across the lab-to-field spectrum [9], as needed to determine the biological and psychological factors related to social behavior or other aspects of vole lifestyle. Behavioral studies have the same requirements as other experiments to be interpretable and advance science: parametric control over a few key independent variables that relate to objective outcome measures. Several groups have designed modified operant tasks or use outdoor enclosures for larger-scale studies of vole bonding. Larger spaces can contain more animals (voles or otherwise), increasing opportunities for interesting decision-making and behavioral manifestation in a way that also enables video tracking and/or neural recordings for studies of social behavior [10] [11] [12] [13]●. But larger enclosures are not necessarily any more or less natural than a benchtop three-chamber assay as the essence of the study is still the same: documenting spontaneous behaviors of small cohorts of animals interacting in a defined spatio-temporal context. Animals presumably do not know they participating in a lab experiment rather than in the wild. If the term 'natural' is applied to these studies, that term might just mean 'bigger enclosure with dirt and temperature variations', but in a less controlled way than if this was custom built in a lab.

For many studies it is desirable or necessary to have lower dimensional behavioral models for testing specific mechanisms or hypotheses. For example, although it is not at all clear what natural behavior is modeled by threat avoidance conditioning (e.g., via pairing tones with foot shocks), the robustness of the system and the expansive literature makes this an ideal paradigm, for example to examine cell-type specific responses or validating a new molecular reagent or optical tool.

**The natural statistics of behavior**

Krakauer et al. [1] provide an ethological perspective on behavior, writing: "correctly labeling something as behavior is contingent on the outcome of an investigation into what the animal does to ensure its survival in its native habitat." This is similar to the thesis of a recent review by Michael Yartsev [14]. So, what are the native habitats and natural behaviors for survival of two of the central species used for biomedical studies: humans and lab mice?

I'm not sure there is such a thing as the native habitat or natural behavior of humans. Is sending a robot to Mars or participating in an online seminar a natural behavior? I think these are both interesting and important things we do as a species. In his wonderful and scholarly book "Natural Neuroscience", Nachum Ulanovsky [15]●● eloquently describes many classic and contemporary studies in systems neuroscience. Figure 1.2 plots different behavioral studies on axes of artificial vs natural and controllable vs uncontrollable. Near the top of artificial behavior is 'Working memory studies in humans,' e.g., memorizing lists of words or images. However, most children in grade school memorize lists or tables for multiplication, grammar, or geography; pre-med students studying for the MCAT must remember large amounts of data. We learned a lot about capacities, mechanisms, and failure modes from such 'artificial' studies [16] [17] [18] [19] [20], analogous to the ear mold studies of Hofman et al. [2] in **Figure 1**. If many individuals are doing this throughout life, and this behavior is something important to our success as a species, is it really 'non-natural'?

For common strains of house mice used for biological studies beyond neuroscience such as C57BL/6 mice, I would argue that the native habitat now has become the institutional vivarium. A vivarium is not mouse cold storage, but rather quite rich and dynamic, enabling animals to interact and observe each other over very long periods. These experiences and the vagaries of specific environments (indoors or outdoors) can have profound consequences on animal state and behavior [21] [22]●. Most species including humans and mice must be behavioral generalists and adapt to the environments they happen to live within. Habitats and ecological niches are not static but incredibly dynamic in terms of threats to survival by temperature and weather, predation, and infection [23, 24] [25] [26]. Species lacking neuro-behavioral capacities to survive these dynamics and adapt to changing conditions will likely go extinct, thus evolutionarily ensuring that existing species can learn and have immense behavioral plasticity on short timescales.

It is more critical to ask not what animals do in 'normal' environments, but what can they do when challenged by unexpected conditions, i.e., in response to something 'abnormal'. One way to quantify 'naturalness' of behavior is to document statistical likelihood of behavioral manifestation in a given social-physical environment, analogous to quantifying likelihood of specific stimulus features such as oriented bars in natural scenes to understand visual processing by the brain [27] [28]. Natural behaviors would be those that are more common, and abnormal behaviors are statistically unlikely given a certain context or set of circumstances. But as with natural scenes, some of the rarest events might be the most interesting and informative towards understanding future behavioral patterns. For example, childbirth is a very rare event in terms of the full lifelong behavioral ethogram, but is extremely important in shaping the rest of an individual's life.

It is also important to have objective outcome measures or quality of life impact. The DSM-V is a statistical descriptor for human behavior. The major challenge for attempts to model psychiatric disorders in non-human species is that we lack a statistical understanding of behavioral likelihoods and outcome impact for anything besides humans. Even in the case of most major disorders, the most likely behaviors for any given moment would be indistinguishable between the psychiatric and neurotypical conditions (e.g., sleeping, eating, walking around). This means that statistical likelihood of a behavior in a given environment (i.e., one definition of 'naturalness') might be necessary but insufficient for understanding and treating these disorders. Instead, rarer events might be much more informative but by nature harder to experimentally document and analyze.

This is now changing, due to massive increases in storage capacities and development of inference tools for analyzing large-scale video data. Recent work by Bedbrook et al. [29]●● is technically masterful, making video recordings of the full lifespan of African killifish for months (**Fig. 3A**). They quantified behavioral statistics (of the kind provided by top-down video, e.g., size, motion, foraging) related to longer or shorter lifespans across a total of 81 continuously-monitored individual animals. They found that size and feeding were not predictive of longer lifespan, but instead greater periods of inactivity earlier in life predicted shorter lifespan overall (**Fig. 3B**).

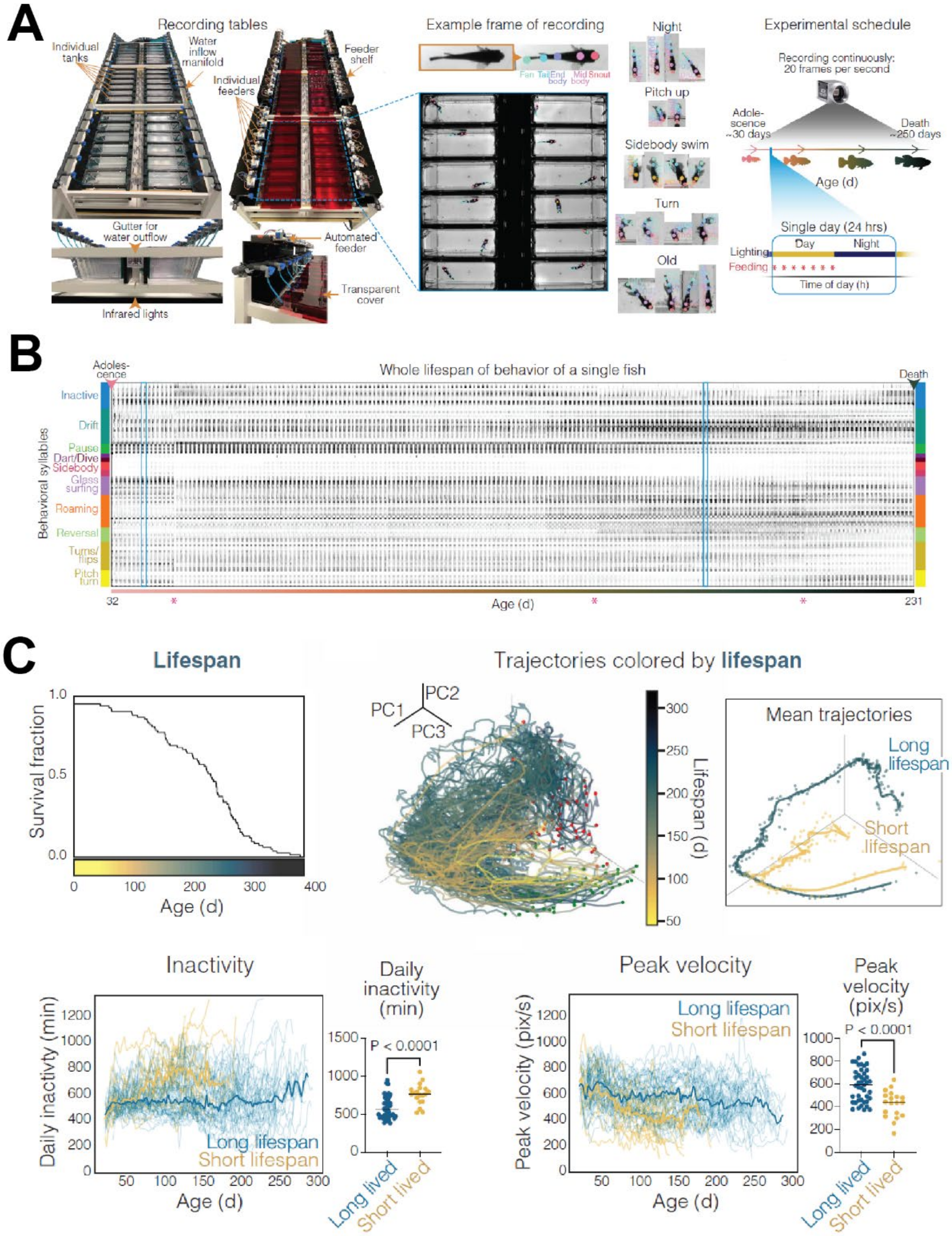

**Figure 3.** Monitoring animal behavior over the lifespan. **A**, Left, 2.8-liter tanks for monitoring killifish from adolescence until death. Middle, imaging and keypoint tracking. Right, timeframe. **B**, Lifespan behavioral syllable usage for single fish from 32 days until death at 231 days. **C**, Top, Kaplan-Meier lifespan curve of N = 81 killifish. Aging trajectories colored by future lifespan. Bottom, correlation between peak velocity and lifespan (ρ = 0.35). From [29]●●.

What options are available for relating moment-by-moment behavior or latent states to end goals? Death is an intuitive and objective outcome measure that can be used to determine the utility and consequences of different behaviors or strategies. However, given its finality and general remoteness in time for most behaviors made by an individual, it is difficult in practice to assign mortality of specific organisms to choices made throughout individual life. These studies remain largely correlative, requiring large sample sizes and focusing on population health measures and lifestyle trends such as caloric intake or distance traveled rather than single consequential behavioral choices of the kind amenable to neural recording and perturbation experiments.

This is why I appreciate that Berendzen et al. [7] documented pup survival (and to some degree adult survival) in their knockout voles (**Fig. 2C**). Infant survival in mammals requires intense and prolonged behavioral engagement by caretakers, on a circumscribed timeframe, localized in space, with outcomes resolving over minutes to days (rather than weeks to years). Questions of behavioral strategies, social interactions, environmental variables, and neural systems can be more adequately addressed with a clear outcome measure at the foundation of species survival.

**So, what behavior?**

Of the four exemplar behavioral systems offered by Krakauer et al. [1], two of them have origins in ethology (jamming avoidance in electric fish, sound localization in owls) while the other two clearly aim to understand and improve something about the human condition (primate motor learning, Parkinsons symptomology). This is one of the main choices when it comes to behavior:

translational research modeling aspects of the human condition, or foundational research (formerly known as basic research) based more or less on subjective criteria (personal interest in a topic).

Effective translational studies require the identification of specific mechanisms that might emulate some part of the human experience. Along with that, a healthy skepticism and willingness to move on to new methods or systems is required, when the limits of the current approach have been reached. This is why it will remain necessary to perform experimental studies in non-human primates, as at some point the physiology or cognitive capacity of other species will fail to adequately approximate that of humans. Optimally, the outcome measures should be interpretable in terms of health and survival of the individual and other members of their family or species.

As for more exploratory foundational science, why not work on whatever we find interesting, mysterious, or exciting? After all, we must communicate the value of our work if others are to continue it after we are done. If an animal can do something (regardless as to the behavioral rarity or particular species), there are neural circuits for these actions which can be examined based on availability of appropriate tools and methods, or the will to generate these tools. A favorite recent example is the Kronauer lab's impressive creation of transgenic GCaMP-expressing ants [30] and a clonal raider ant reference brain [31]●●. Cognitive functions and survival skills might only manifest under duress or under severe circumstances. As Krakauer et al. [1] emphasize, building a theoretical framework to quantify and predict dynamics and criticalities is essential, but perhaps is only really achievable after limit conditions have been examined through training or testing animals under highly artificial or non-natural conditions.

## Acknowledgments

I thank D. Aronov, C.M. Constantinople, Z.R. Donaldson, D.S. Manoli, and M.A. Long for comments, and grants from the Human Frontier Science Program and the NIH (HD088411, NS107616, and NS138066) for funding.